\documentclass[aps,prd,floats,superscriptaddress]{revtex4}
\usepackage{graphicx}
\newcommand{\newc}{\newcommand}
\newc{\beq}{\begin{equation}}
\newc{\eeq}{\end{equation}}
\newc{\beqa}{\begin{eqnarray}}
\newc{\eeqa}{\end{eqnarray}}
\newc{\bray}{\begin{array}}
\newc{\eray}{\end{array}}
\newc{\IM}{\mbox{\sl{Im}}}
\newc{\RE}{\mbox{\sl{Re}}}

\newc{\nonr}{\nonumber}
\newc{\hs}{\hskip 3mm}
\newc{\ra}{\rightarrow}
\newc{\TR}{\mbox{\sl{Tr}}}
\newc{\tri}{\triangle}

\begin{document}

\pagestyle{plain}

\title{Electric Dipole Moment in the Split Supersymmetry Models}

\author{Darwin Chang}
\email{chang@phys.nthu.edu.tw}
\affiliation{Physics Department, National TsingHua University, Hsinchu  300, Taiwan}

\author{We-Fu Chang}
\email{wfchang@phys.sinica.edu.tw}
\affiliation{Institute of Physics, Academia Sinica, Taipei 115, Taiwan}

\author{Wai-Yee Keung}
\affiliation{Physics Department,
University of Illinois,
Chicago, IL 60607-7059 }
\date{\today}

\begin{abstract}
We study an important contribution to the electric dipole moment (EDM)
of the electron (or quarks) at the two-loop level due to the $W$-EDM
in the recently proposed scenario of split supersymmetry.  This
contribution is independent of the Higgs mass, and it can enhance the
previous estimation of the electron (neutron) EDM by $20-50\%$
($40-90\%$). Our  formula is  new  in its analytical form.

\end{abstract} \pacs{Who cares?}
\maketitle
\section{Introduction}
Supersymmetry (SUSY) has been  considered an extended  symmetry
beyond Standard Model (SM) to solve the gauge hierarchy problem,
to stabilize the scalar sector,
and to provide a theoretical ground for possible unification of gravity with
all other fundamental forces.
However, Minimum Supersymetrically  Standard Model (MSSM)
predicts plethora of new superpartner particles, but
none of them has been observed yet.
Therefore soft terms are introduced to break  SUSY
but keep the feature of taming the quadratic divergence.
Unfortunately, it is well known that
at the electroweak (EW) scale the softly broken SUSY
generates many unwanted phenomenological problems,
such as  flavor changing neutral currents, CP violation, and so on.
Motivated by the string landscape scenario and the cosmological
constant problem,  Arkani-Hamed and Dimopoulos have recently
proposed a scenario \cite{Arkani-Hamed:2004fb}
(dubbed Split SUSY) that SUSY is  broken at an energy scale
way beyond the collider search and could be even near
the scale of the grand unification theory (GUT).
As a result the scalar superpartners of Standard Model fermions are all super heavy.
On the other hand, fermions are protected by symmetries,
such as, to be more specifically,
chiral symmetry, R-symmetry and PQ symmetry,
so they acquire masses around electroweak to TeV or so.
By doing so, the phenomenological problems of SUSY at the EW scale
can be avoided.
However, the existence of a  CP-even SM like Higgs with mass around
$100-250$ GeV requires  a fine tuning in the Higgs potential.
To address this fine tuning problem, they
argue that extreme fine turning  is required for solving
the cosmological constant
problem which is viewed as choosing a stringy ground state
with small cosmological constant from an extreme  huge  pool of vacuum
candidates.
Admitting  this kind of fine tuning, one  is no longer worried about
the   naturalness principle and the fine turning in the Higgs sector.
(Recently it was pointed out by \cite{Drees:2005cp} that
extra fine tuning is needed to get  a reasonable values for $\tan\beta$.)

Phenomenologically, the characteristics of the Split SUSY can be summarized
as following.
(1) All scalars, except the CP-even SM like Higgs, are super heavy
$\sim 10^9$ GeV - $M_{\rm GUT}$.
(2) Gaugino  masses are around the EW scale  to TeV protected by
R-symmetry (PQ symmetry).
(3) $\mu$ parameter is around the EW scale such that the lightest neutralino
can annihilate effectively to give the dark matter density.
(4) Coupling unification still works, mainly due to the gauginos contributions.

There are already many discussions on detecting or testing the split
SUSY by accessible
 and/or near future
experiments \cite{Arkani-Hamed:2004yi,generalPH,NormalHM, HeavyHM},
 on the connection with the neutrino masses \cite{nutrino_mass},
and  on the
implication in cosmology and astrophysics\cite{astrophysics}.
In this note we will only  focus on the inherent contributions to EDM  in the Split
SUSY scenario.

\begin{figure}[ht]
  \centering
    \includegraphics[width=0.45\textwidth]{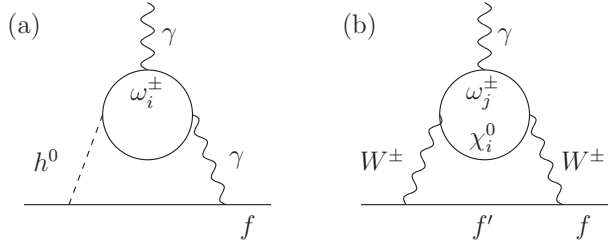}
  \caption{The  2 loop diagrams contribute to fermion EDM. }
  \label{fig:2loop}
\end{figure}

In split SUSY,
the gaugino masses parameters,
$M_{1,2,3}$ for $U(1), SU(2)$ and $SU(3)$ gauge group respectively,
as well as the Higgsino mixing mass parameter $\mu$,
are all around the EW scale.
Consequently, the charginos and neutralinos have masses at the same
scale.  These parameters are generally complex with respect to each
other, and their mutual phases cannot be removed by redefinition of
fields. If so, the CP violation in the gaugino-Higgsino sector is genuine,
and it can give rise to the EDM of an elementary particle at low energy.
Nevertheless, all possible one-loop contributions to EDM are highly
suppressed by the super heavy scalar mass in the loop.
Thus the leading sources of the EDM starts at the two-loop level where
SM particles and EW charginos and neutralinos run in the loops.  A study of
possible EDM due to the complex pseudo-scalar coupling of light
neutral Higgs, see Fig.~\ref{fig:2loop}(a), has been done in
\cite{Arkani-Hamed:2004yi, Chang:2002ex}.  This type of EDM will be referred as
$d^{h^0}$.  However, we want to point out that the $W$ gauge boson EDM, see
Fig.~\ref{fig:2loop}(b), can give same order of magnitude contribution
to EDM of SM fermion, denoted as  $d^W$.
Moreover, this EDM contribution, $d^W$,
is independent of the Higgs mass.
In addition, the two diagrams actually
depend on different combinations of CP violating
phases and therefore, in some occasions, $d^W$ still
contributes even when the CP violation effect in $d^{h^0}$ vanishes
accidentally.
Note that  the two types of contributions are totally determined by
7 free parameters: $\tan\beta$,
three mass parameters, $( M_1, M_2, \mu)$, and three arbitrary CP phases
$( \phi_1, \phi_2, \phi_\mu )$.

In MSSM, a similar 2-loop $H^\pm W^\mp\gamma$ diagram
with one of the $W^\pm$ lines in Fig.~\ref{fig:2loop}(b)
substituted by $H^\pm$ turns out to give more important EDM contribution
than $d^W$ for a large range of the
parameter space\cite{in_preparation}.
But such contribution vanishes in the split SUSY due to
the decoupled super heavy charged Higgs.

We will carefully study the resulting EDM from the combined two contributions,
mentioned above, within a reasonable parameter space.

\section{2-loop EDM}
We start from the  relevant Lagrangian which could lead to EDM.
It can be written in a general form  as
\beq
\label{eq:Lint}
{\cal L} \supset  +{ g \over \sqrt{2}} \overline{\omega_j^+}\gamma^\mu
[O^L_{ij}P_L + O^R_{ij} P_R] \chi_i^0  W_\mu^+
-{g \over \sqrt{2}} O'_i \overline{\omega_{iR}^-} \omega_{iL}^-
h^0 + h.c.
\eeq
The couplings are
\beqa
O^R_{ij}&=&\sqrt{2} N_{2i}^* C^L_{1j} + N_{3i}^* C^L_{2j}\,,\;
O^L_{ij}=\sqrt{2} N_{2i} C^R_{1j} - N_{4i} C^R_{2j}\,,\\
O'_i &=& (C^R_{1i})^*C^L_{2i} \cos\beta
 +(C^R_{2i})^* C^L_{1i} \sin\beta \ ,
\eeqa
where  the unitary matrices $C^{L,R}$ and $N$ are defined to
diagonalize the chargino and neutralino mass matrices
\beqa
{\cal M}_C &=& \left( \begin{array}{cc}
M_2 e^{i\phi_2} & \sqrt{2} M_W c_\beta \\
\sqrt{2} M_W s_\beta & \mu e^{i \phi_\mu} \end{array}\right)\,, \\
{\cal M}_N &=& \left( \begin{array}{cccc}
M_1 e^{i\phi_1}&0 & - M_Z s_W c_\beta &  M_Z s_W s_\beta\\
0& M_2 e^{i\phi_2} &  M_Z c_W c_\beta & - M_Z c_W s_\beta\\
- M_Z s_W c_\beta &  M_Z c_W c_\beta &0 &  -\mu e^{i\phi_\mu}\\
 M_Z s_W s_\beta &  -M_Z c_W s_\beta &  -\mu e^{i\phi_\mu}&0
\end{array}\right)   \ ,
\eeqa
with  $ C^{R\dag}{\cal M}_C C^L=diag\{ m_{\omega_1}, m_{\omega_2}\}$
and $N^T {\cal M}_N N=diag\{ m_{\chi_1}, m_{\chi_2},m_{\chi_3},m_{\chi_4} \}$.
The diagonalized masses are positive and real. We use the convention that
$m_{\omega_1}< m_{\omega_2}$ and
$m_{\chi_1}< m_{\chi_2}<m_{\chi_3}<m_{\chi_4}$.
Notation $s_W(s_\beta)$ stands for $\sin\theta_W (\sin\beta)$ and
$\tan\beta=v_u /v_d$.
The matrices $C^{L,R}$ are not uniquely defined. However the resulting EDM is
basis independent.

These two-loop diagrams or similar ones  have been calculated many times in the
literature\cite{Chang:2002ex,in_preparation,Barr:1990vd,
Chang:1998uc,Bowser-Chao:1997bb,Pilaftsis:2002fe}.
Here we just summarize and report the essential results for the most important contribution
from two gauge invariant subsets.

In Fig.~\ref{fig:2loop}(a),
since the coupling between the  SM Higgs and charged fermion
is pure scalar like, only the pseudo-scalar form factor
of the photon-photon-Higgs  vertex in the upper loop will contribute to EDM.
We denote the momenta and polarizations for the photon-photon-Higgs vertex
as $h^0(p=q+k)\ra \gamma(k,\mu)+\gamma(q,\nu)$. In this way,
the pseudo-scalar part from the  can be derived to be:
\beq
i\Gamma^{\mu,\nu}= i {g^2 e \over 4\sqrt{2}\pi^2 }
\sum_{i=1}^2\hbox{ Im }O'_i \; m_{\omega_i}
\int^1_0 d\gamma {1-\gamma \over m_{\omega_i}^2-\gamma(1-\gamma)p^2  }
\times \epsilon^{\mu,\nu, \rho,\lambda}\, k_\rho\, q_\lambda\,.
\eeq
This form factor  is further connected  to the SM charged fermion line
and the resultant EDM becomes
\beq
{d^{h^0}_f \over e} = {Q_f \alpha^2 m_e \over 4\sqrt{2}\pi^2 M_H^2 s_W^2}
\sum_{i=1}^2\hbox{ Im }O'_i {m_{\omega_i} \over M_W }
{\cal F}\left(\frac{m_{\omega_i}^2}{M_H^2}\right)   \ ,
\eeq
where $Q_f$ is the charge of SM fermion $f$ and the function ${\cal F}$ is
\beqa
{\cal F}(x)&=& \int^1_0 d\gamma {(1-\gamma) \over
x-\gamma(1-\gamma)}\ln{x\over \gamma(1-\gamma)}\nonr\\
&=&\mbox{Re}\left\{
 {1\over\sqrt{1-4x}}\left[\ln x \ln{\sqrt{1-4x}-1  \over \sqrt{1-4x}+1 }
+Li_2\left({2\over 1-\sqrt{1-4x} }\right)
-Li_2\left({2\over 1+\sqrt{1-4x} }\right)
\right]\right\}\,.
\eeqa
The result agrees with the analysis of
\cite{Arkani-Hamed:2004yi} when $x>1/4$.
However, we emphasize that only the real part is taken when $x<1/4$
because the imaginary part is only a mathematical artifact.
Note that $Li_2(z)=-\int_0^z \ln(1-t) (dt/t)$.

The diagonalization of the $2\times 2$ chargino mass matrix and
the coupling $O'_i$
can be done analytically, see for example\cite{Chang:2002ex}.
We note by passing that this EDM  is
proportional to the $\hbox{ Im }(\mu M_2)$. In other word, $d^{h^0}$
vanishes in  the parameter space where $\arg(\mu M_2)=0\, \mbox{mod}\, 2\pi$.

To calculate the $d^W$, we can first integrate out the upper loop
in Fig.~\ref{fig:2loop}(b).
Following  \cite{Hagiwara:1986vm}, the CP
violating form factor $f_6$ for $W^+(p=q+k,\nu)\ra W^+(q,\lambda)+\gamma(k,\mu)$
is defined by the effective vertex
\beq
\label{eq:CPXF6}
i \Gamma^{\mu, \nu,\lambda} = -i f_6 \epsilon^{\rho,\mu,\nu,\lambda}
k_\rho\, .
\eeq
Another parameterization of this  vertex
can be found in \cite{Marciano:1986eh}, where
its implication to the electron EDM was studied with a short-distance
cutoff.
The cutoff is unnecessary because
the general interaction of  Eq.~(\ref{eq:Lint}) prescribes  the $q^2$
dependence in $f_6$,
\beq
f_6(q^2) ={e \alpha \over 2\pi s^2_W} \sum_{i=1}^4 \sum_{j=1}^2
\hbox{ Im }(O^L_{ij} O^{R*}_{ij})
\int^1_0 d\gamma {m_{\chi_i} m_{\omega_j} (1-\gamma)\over (1-\gamma)
 m_{\omega_j}^2 + \gamma m_{\chi_i}^2 -\gamma(1-\gamma)q^2}       \ ,
\eeq
which agrees with \cite{Chang:1990fp,West:1993tk, Kadoyoshi:1996bc}.
The resulting EDM of charged fermion $f$ due to $f_6$ is finite,
\beqa
\label{eq:WEDM}
{d^W_f\over e} &=&  \pm {\alpha^2 m_f \over  8 \pi^2 s^4_W M_W^2}
 \sum_{i=1}^4 \sum_{j=1}^2  {m_{\chi_i}m_{\omega_j}
\over M_W^2} \hbox{ Im }(O^L_{ij} O^{R*}_{ij})
 {\cal G}\left( r^0_i, r^\pm_j, r_{f'} \right)\,,\\
 {\cal G}\left( r^0_i, r^\pm_j, r_{f'} \right)
 &=& \int^\infty_0 dz\int^1_0 { d\gamma \over \gamma} \int^1_0  dy\;
{y\, z\, (y +z/2 )\over (z+R)^3(z+K_{ij})}\nonr\\
&=& \int^1_0 { d\gamma \over \gamma} \int^1_0 dy\, y  \left[
{(R-3K_{ij})R+2(K_{ij}+R)y \over 4 R(K_{ij}-R)^2 }+{K_{ij}(K_{ij}-2y)
\over 2(K_{ij}-R)^3}\ln\frac{K_{ij}}{R}
 \right]\,.
\eeqa
The plus(minus) sign in front the right-handed side of Eq.(\ref{eq:WEDM})
corresponds to the fermion $f$ with weak isospin
$+(-)1/2$.
The short-hand symbols  $K,R, r$s are defined as
\beq
R=y+(1-y)r_{f'}\,,\;
K_{ij}= {r^0_i \over 1-\gamma}+{r^\pm_j \over \gamma}
\,,\;
r^\pm_j \equiv {m_{\omega_j}^2 \over M_W^2}\,,\;
r^0_i \equiv {m_{\chi_i}^2 \over M_W^2}\,,\;
r_{f'} \equiv {m_{f'}^2 \over M_W^2}  \ .
\eeq
Here  $f'$ is the electroweak $SU(2)$ partner of  $f$.
In the large $K_{ij}$ limit, the leading expansion result agrees
with \cite{Kadoyoshi:1996bc}.
However we emphasize  that our Eq.(\ref{eq:WEDM}) is an
exact formula which does not appear previously. For example,
our result is numerically few percents larger than those given in
\cite{West:1993tk,Kadoyoshi:1996bc}
when the  chargino and neutralino masses are around the EW scale.

We diagonize the $4\times4$ neutralino mass matrix directly by the numerical method.

\section{Numerical results}
\begin{figure}[ht]
  \centering
   \includegraphics[width=0.6\textwidth]{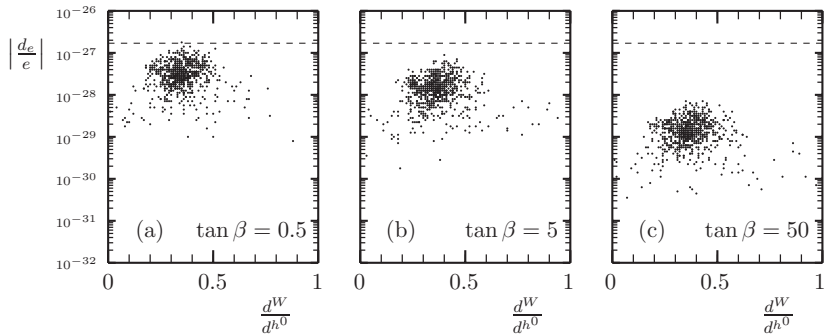}
  \caption{The  total EDM and the ratio of $d^{W}/d^{h^0}$. }
   \label{fig:eEDM_ratio}
\end{figure}

In previous  study\cite{Arkani-Hamed:2004yi,Chang:2002ex},
the EDM $d^{h^0}$ was shown as the
function  of the ratio of $m_{\omega_2}/m_{\omega_1}$. However
a general scheme is represented by any points in a space of seven
parameters mentioned above.
Therefore we evaluate both $d^{h^0}$ and $d^W$ by randomly scanning
the following parameter space,
$200$~GeV $< M_1, M_2, \mu <1.0$ TeV,
$120$~GeV $<M_H <170$~GeV and all the three CP phases vary
within $[0, 2\pi]$.
The above range of the Higgs mass
was suggested by \cite{NormalHM}.
However, some variants allow the light Higgs to be as heavy as
400~GeV\cite{HeavyHM}.
The numerical result of five hundred randomly selected points
are shown in Fig. \ref{fig:eEDM_ratio} for $\tan\beta= 0.5,
5.0$, and $50$ respectively.
The current upper limit on the electron EDM,
$<1.7\times 10^{-27}$ e-cm $95\%$ CL\cite{eEDMexp},
is shown as the dash line in the graphs.

From these plots,
we notice that  for electron:

(1) contributions of  $d^{h^0}$ and $d^W$ share
the same sign and the ratio $d^W/d^{h^0}$ lies around
$0.2-0.5$ for a light Higgs mass within $120 - 170$ GeV.
Indeed there are very few points within the scanned range not appear
in the plots. For those rare
cases, the reason can be identified as
$\arg(\mu M_2)\sim 0$ or $d^{h^0} \ll d^W$ and the
total possibility is less than $1\%$.
(2) The electron EDM is around $10^{-28}-10^{-27}$ e-cm
for $\tan\beta=0.5$ and it decreases to  $10^{-30.5}-10^{-29.5}$ e-cm
when $\tan\beta=50$.

\begin{figure}[ht]
  \centering
    \includegraphics[width=0.48\textwidth]{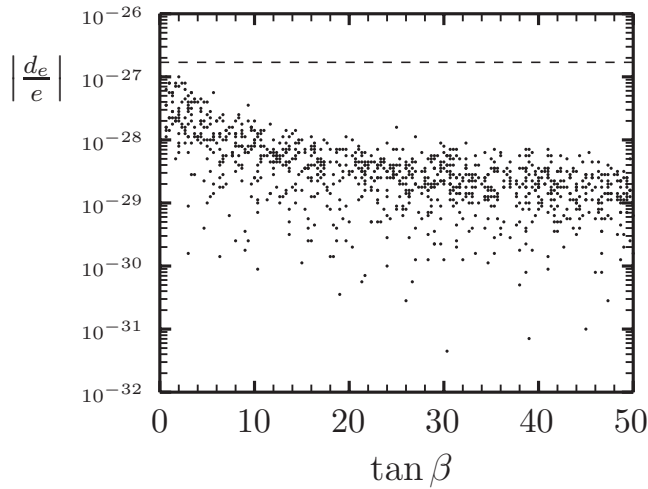}
  \caption{The electron EDM v.s $\tan\beta$. }
  \label{fig:e_EDM_all}
\end{figure}

The electron EDM versus $\tan\beta$ is shown in
Fig.~\ref{fig:e_EDM_all}. Based on the parameter scan, it seems very
promising in the observation of the electron EDM by  experiments
with the sensitivity of $10^{-29}$ e-cm \cite{NewEEDM}.

\begin{figure}[ht]
  \centering
   \includegraphics[width=0.6\textwidth]{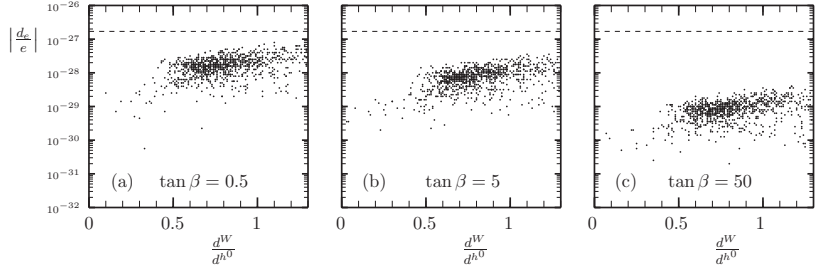}
  \caption{The  total EDM and the ratio of $d^{W}/d^{h^0}$ for $400GeV <M_H<600 GeV$. }
   \label{fig:eEDM_ratioHH}
\end{figure}

\begin{figure}[ht]
  \centering
    \includegraphics[width=0.48\textwidth]{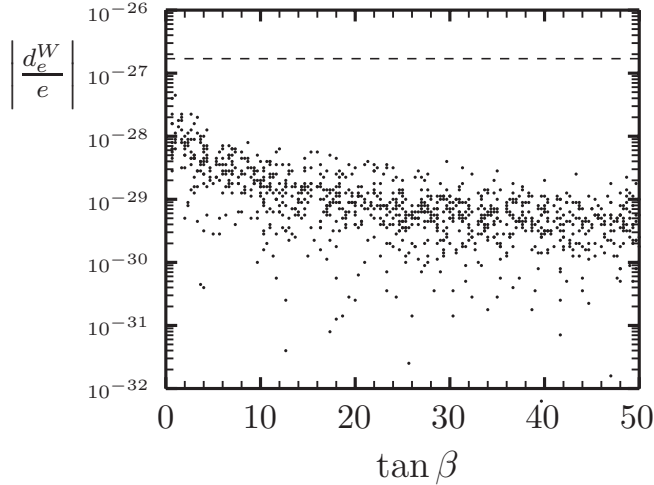}
  \caption{The electron EDM from $d^W$ alone v.s $\tan\beta$. }
  \label{fig:e_EDM_all_W}
\end{figure}

As the lightest neutral Higgs becomes heavier, the $d^W$ contribution
to the EDM of the charged SM fermion turns out to be increasingly
important.  The values of $d^{h^0}$ and $d^W$ are roughly compatible when
$M_H \sim 600$ GeV, see Fig.~\ref{fig:eEDM_ratioHH}, and $d^W$
dominates over $d^{h^0}$ for  larger  $M_H$.
In the extreme case of a super heavy Higgs,  $d^W$ is the sole
contribution to the EDM of SM fermions. In Fig.~\ref{fig:e_EDM_all_W}
we show the only contribution $d^W$ in  this extreme limit, where
the EDM is roughly half order of magnitude smaller than that of a
light Higgs mass within $120-170$ GeV already illustrated in
Fig.~\ref{fig:e_EDM_all}.   Nevertheless, an electron EDM around
$10^{-29}$e-cm predicted in the extreme case of a super heavy Higgs is
still probably detectable in the future experiment.

In the split SUSY models,
the charged lepton EDMs  follow the simple mass scaling law and
the muon EDM is  given by
the electron EDM scaled up by the factor of $m_\mu/m_e$,
which is quite different  from some models, for example
see\cite{Mscal}. Therefore, models can be  distinguished
by comparing the  electron and the muon EDMs.
However, split SUSY predicts
the $d_\mu$ to be roughly $10^{-24.5}- 10^{-27}$e cm,
which is 6 to 7 orders of magnitude lower
than the current limit\cite{muedm} and it will be a great
challenge for the newly proposed $d_\mu$ measurement\cite{NewMuEDM}.

Now we turn our attention to the neutron EDM.

In MSSM, usually the chromo dipole moment is the dominant contribution
to the neutron EDM due to the large $\alpha_s$ of the strong interaction.
However, in the split SUSY models, the CP phases associated with
gluinos can always be shuffled off upon the squarks mass matrix by
phase redefining of the gluino field.  The chromo dipole moment
therefore vanishes because all the squarks are decoupled from the low
energy physics and $d^{h^0}$ and $d^W$ become the leading contribution to
the neutron EDM.

Given the nonperturbative nature of hadron physics, it is not clear how to make
reliable theoretical prediction on the neutron EDM under control.
However, as an order of magnitude estimation, the quark model prediction
$d_n =( 4 d_u -d_d)/3$ can be used to give a rough estimation of the
neutron EDM.  By trivially scaling up the fermion masses and replacing
the fermion charge accordingly, we can express the neutron EDM as
\beq
\label{eq:quarkM}
d^{h^0}_n = -\left({8 m_u +m_d \over 9 m_e}\right) d_e^{h^0}\,,\;
d^W_n = -\left({4 m_u +m_d \over 3 m_e}\right) d_e^W\,.
\eeq
In arriving at the last expression of $d^W_n$, we have ignored
masses of $SU(2)$ doublet partners, the  $u$ and $d$ quarks, in the
loop.
Since the light quark masses are much less than $M_W$, this approximation is
quite safe.

As in the electron case, the relative sign  between
the two contributions is also positive.

\begin{figure}[ht]
  \centering
    \includegraphics[width=0.5\textwidth]{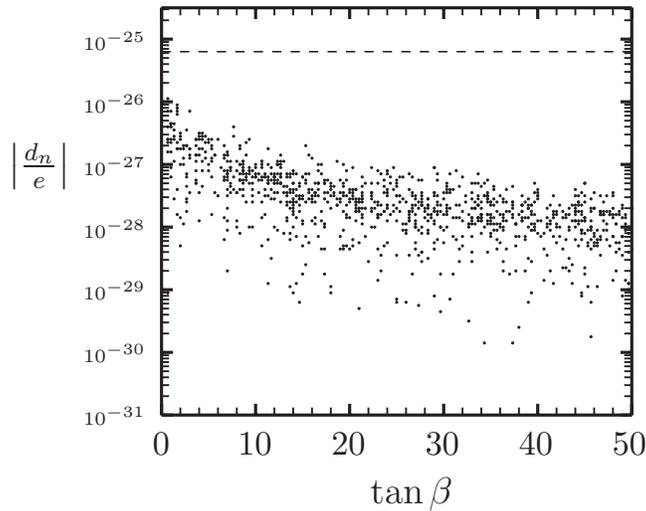}
  \caption{The neutron EDM v.s $\tan\beta$. }
  \label{fig:n_EDM_all}
\end{figure}

The estimation of the resulting neutron EDM
is displayed in Fig.~\ref{fig:n_EDM_all},
where the current quark masses, $m_u=3$ MeV and $m_d=6$ MeV,
have been used.  Here, the same range  of the parameter space
is scanned as in Fig.\ref{fig:e_EDM_all}.
Note that the proper choice of  quark masses is still controversial,
however such a question is beyond the scope of this article.
The readers should keep in mind that our estimation of
neutron EDM is conservative and the prediction
could receive substantial enhancement due to
the unknown nature of hadronic physics.

The current upper limit of neutron EDM, $<6.3\times 10^{-26}$ e-cm at $90\%$
CL\cite{PDG}, is also displayed as the dash line in the graph.
In comparison with the electron EDM, the $d^W$ becomes more important
in the neutron EDM study due to the naive enhancing factor in
Eq.(\ref{eq:quarkM}).
While the sum of  $d^W$ and $d^{h^0}$ is still below the current upper limit,
$d^W$  plays an indispensable role in the neutron EDM.

\section{Conclusion}
This article studies the EDM in the scenario of Split
SUSY.  We point out that an overlooked  but important  two-loop contribution,
Fig.~\ref{fig:2loop}(b), due to the $W$ gauge boson EDM,
has to be included together with others given by  previous EDM
study\cite{Arkani-Hamed:2004yi}
where only Fig.~\ref{fig:2loop}(a) type diagram was considered.

(1) For most of the parameter space, the $W$-EDM diagram enhances the
previous estimation of the electron EDM by $20- 50\%$.

(2) For some special circumstances  that $\arg (M_2 \mu)\sim 0$ or
the neutral Higgs are super heavy, the EDM contribution from
$d^{h^0}$ vanishes, and the fermion EDM will be dominated by $d^W$.

(3) Combining these two EDM contributions, we have scanned the whole
parameter space and found that the electron EDM is likely to be seen
in next run of EDM experiments.

(4) We estimate the  neutron EDM by using the naive quark model.
With typical current quark masses $m_u=3$ MeV and $m_d= 6$ MeV,
the numerical result indicates that the contribution from $d^W$ is
about the same size of $d^{h^0}$. However, the overall result is about
an order of magnitude or more below the current experimental bound.

\end{document}